\documentclass[prl,reprint,nobibnotes]{revtex4-1}

\usepackage{color}

\usepackage{graphicx}
\usepackage{booktabs}
\usepackage{bm}
\usepackage{longtable}
\usepackage{amsmath}
\usepackage{dcolumn}
\usepackage{placeins}
\usepackage{expdlist}

\usepackage{siunitx}
\usepackage{ulem}

\begin{document}

\bibliographystyle{apsrev}

\title {Indirect pumping of alkali-metal gases in a miniature silicon-wafer cell}

\author{J.D.\, Zipfel}
\email{The author to whom correspondence may be addressed: jake.zipfel@npl.co.uk}
\affiliation{National Physical Laboratory, Hampton Road, Teddington TW11 0LW, United Kingdom}
\author{G.\, Quick}
\affiliation{INEX Microtechnology Ltd., Herschel Annex, Kings Road, Newcastle upon Tyne, NE1 7RU, United Kingdom}
\author{B.\, Steele}
\affiliation{INEX Microtechnology Ltd., Herschel Annex, Kings Road, Newcastle upon Tyne, NE1 7RU, United Kingdom}
\author{P. Bevington} 
\affiliation{National Physical Laboratory, Hampton Road, Teddington TW11 0LW, United Kingdom}
\author{L.\, Wright}
\affiliation{National Physical Laboratory, Hampton Road, Teddington TW11 0LW, United Kingdom}
\author{J.\, Nicholson}
\affiliation{School of Physics and Astronomy, University of Birmingham, Edgbaston, Birmingham B15 2TT, United Kingdom}
\author{V.\, Guarrera}
\affiliation{School of Physics and Astronomy, University of Birmingham, Edgbaston, Birmingham B15 2TT, United Kingdom}
\author{W.\, Chalupczak}
\affiliation{National Physical Laboratory, Hampton Road, Teddington TW11 0LW, United Kingdom}

\date{\today}

\begin{abstract}
Atom spin sensors occupy a prominent position in the scenario of quantum technology, as they can combine precise measurements with appealing miniature packages which are crucial for many applications. In this work, we report on the design and realization of miniature silicon-wafer cells, with a double-chamber configuration and integrated heaters. The cells are tested by systematically studying the spin dynamics dependence on the main pump parameters, temperature, and bias magnetic field. The results are benchmarked against cm-sized paraffin-coated cells, which allows for optimisation of operating conditions of a radio-frequency driven atomic magnetometer. In particular, we observe that, when indirect optical pumping is performed on the two cells, an analogous line narrowing mechanism appears in otherwise very different cells' conditions. Competitive results are obtained, with magnetic resonance linewidths of roughly 100 Hz at the maximum signal-to-noise ratio, in a non-zero magnetic field setting, and in an atomic shot-noise limited regime.

\end{abstract}

\maketitle

\section{Introduction} 
In recent years, the combination of technological development and the progress made in understanding underlying operating principles has driven atomic magnetometers and co-magnetometers from lab-based proof-of-principle demonstrations towards a multitude of in-the-field applications. One of the crucial requirements for many real-life implementations, from bio-magnetism detection \cite{Limes2020,Bezsudnova22} to navigation \cite{Walker2016}, is the miniaturisation of the sensor's head. More generally, the reduction of the Size, Weight, Power and Cost (SWaP-C) envelope of these atomic devices can potentially lower operating costs, and consequently enable mass-scale production. In this context, one of the critical components for the miniaturisation of atomic (co)-magnetometers is the vapour cell. 

The first miniature atomic vapour cells, based on a silicon wafer, were realized as a part of the development of Coherent Population Trapping (CPT) clocks \cite{Liew2004, Shah2007, Knappe2007, Kitching2018}. 
The cell volume is enclosed between glass windows which are attached to a silicon wafer via anodic bonding. Since their first realization, these cells have been  built in a variety of configurations, which offer different improvements in performance \cite{Woetzel2011, Knapkiewicz2018}. For example, condensation of the alkali-metal atoms on the cell windows was averted by the use of gold spots mimicking a temperature gradient \cite{Karlen2018}. Extension of the cell lifetime, by reducing the atomic diffusion through the cell walls, has been achieved by coating the inner surfaces with aluminium oxide \cite{Woetzel2013, Karlen2017}. 
Thick wafers, multiple-wafer stacks, or mirrored surfaces inside the cell have been instead used to increase the optical path and thus the number of addressable atoms \cite{Pétremand2012, Dyer2022a, Dyer2022}. 
Finally, in parallel with the development of the cell structure, various methods of alkali-metal deposition have been explored \cite{Knapkiewicz2019}. Due to the novel and specific conditions, many of the engineering solutions introduced in the field of miniature cells also require a complementary understanding of the atomic vapour polarisation dynamics in these environments.     

In this paper we report on the realization of a versatile, cost-effective, mm-sized wafer atomic cell with a double-chamber arrangement and integrated heating circuit. Separation of the alkali-metal storage from the interaction chamber, as well as engineering of a temperature gradient across the cell, prevents reduction of the optical access and the degradation of the spin coherence due to spurious deposition at the cell windows. Moreover, with the use of azide as an alkali-precursor we can greatly simplify the production process, without any sign of compromising the obtained performances.
Our novel cells are thoroughly tested for operation in conditions of indirect pumping, a technique which has proven to be resourceful in paraffin-coated cells \cite{Gartman2015}, but which has not previously demonstrated in miniature wafer-based cells. 

Evacuated paraffin coated cells (without buffer gases) set the metrological benchmark for atomic magnetometers, due to their long coherence lifetimes and the relative simplicity of their pumping dynamics enabled through theoretical description of measured rf spectra. Experimental comparison of the observable dependencies (e.g., linewidth on pump power) can be used to gain direct insight into the pumping dynamics of the more complex scenario of miniature cells filled with inert buffer gas.

\begin{figure}[h!]
\includegraphics[width=\columnwidth]{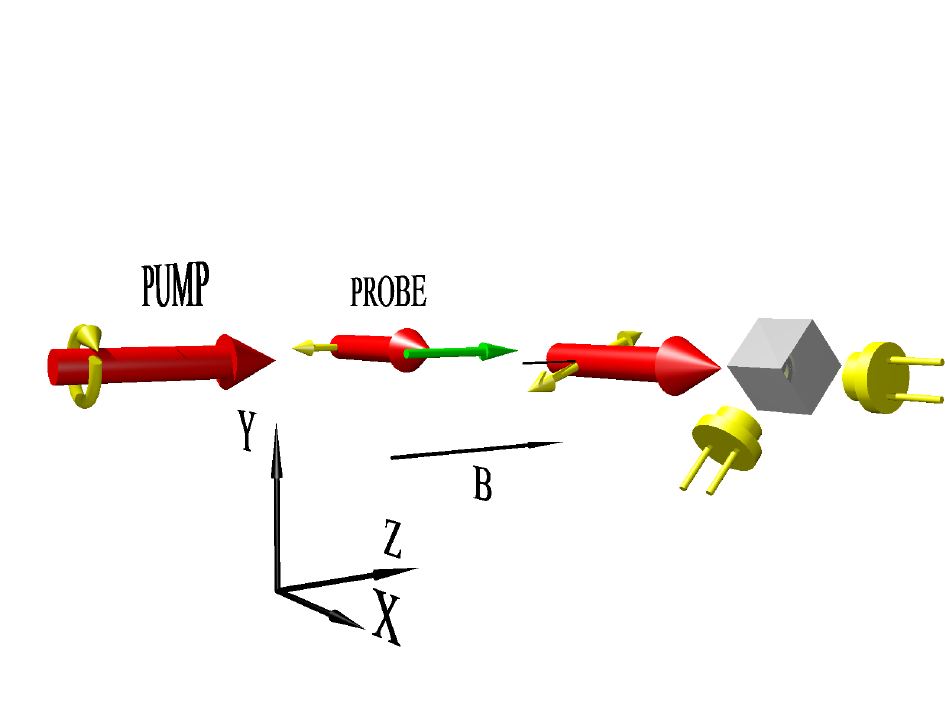}
\caption{
Schematics of the measurement setup. The circularly polarized pump beam propagating along the bias magnetic field direction, $\hat{z}$ axis, creates atomic polarisation along with SEC. The linearly polarized probe, propagating along $\hat{x}$, monitors the atomic coherence precession via the paramagnetic Faraday effect. 
}\label{fig:Setup}
\end{figure}

In particular, we focus on the optimisation of the atomic polarisation and coherence lifetime for atomic magnetometry applications. The measurements are conducted in a pump/probe configuration, where the circularly polarised beam creates a stationary collective atomic spin component, while the linearly polarised probe beam monitors the atomic coherences generated by a radio-frequency (rf) magnetic field, Fig.~\ref{fig:Setup}. The presence of a buffer gas, naturally released by the azide, determines the diffusive character of the thermal atomic motion inside the cell volume \cite{Bevington23}. Different contributions to the atomic polarisation process are identified through the characterisation of the signal’s dependence on pump power, and the temperature dependence enables discrimination between different dynamical regimes dominated by optical pumping or spin-exchange collisions (SEC) of optically polarised atoms. The comparison with analogous measurements taken in a paraffin-coated cell helps identifying the optimal conditions for operation (i.e. pump parameters, and atomic gas temperature), which allow us to obtain competitive performances in the scenario of miniature cells \cite{Lucivero22, Hunter23} with rf resonance widths as good as $\SI{100}{\hertz}$ in non-zero field operation. In particular, we identify a narrowing mechanism in our wafer cells which resembles the one observed in the remarkably different conditions of a paraffin-coated cell, in a SEC-limited regime.   

The paper is structured as follows: first, essential details of the measurement instrumentation are provided. Then, a few aspects of the cell design, manufacturing and characterisation processes are discussed. A brief summary of the measurement method is followed by the analysis of the rf spectrum’s dependence on pump power, recorded at various vapour temperatures, and different magnetic fields.

\section{Experimental setup} 

The measurement setup, schematized in Fig.~\ref{fig:Setup}, is enclosed in a Twinleaf cylindrical shield (MS-1LF) to limit the influence of the ambient magnetic field. The static magnetic bias field is produced by a set of coils within the shields. In separate measurements, we have confirmed that the bias magnetic field inhomogeneity over the cell extension is as good as $0.01\%$, and does not contribute to the profile linewidth. The measurements reported have been performed in a temperature range between $25^{\circ}C$ and $120^{\circ}C$ (atomic densities $n_{\text{Cs}}=1.2 \times10^{11} \text{cm}^{-3}$ to $1.0 \times10^{13} \text{cm}^{-3}$). The cell and heater are enclosed in a 3D printed holder, and the heater is driven by a current sinusoidally modulated at a frequency of 1 MHz. Pumping is performed by a circularly polarised laser beam, frequency tuned close to the $6\,^2$S$_{1/2}$ F=3 $\rightarrow{} 6\,^2$P$_{3/2}$ F'=2 transition (caesium D2 line, $\SI{852}{\nano\meter}$), propagating along the direction of the bias static magnetic field, Fig.~\ref{fig:Setup}. The power of the pump beam is controlled by an acousto-optic modulator operating in double-pass configuration. The atomic coherence is created by coupling the polarized atoms to an rf field generated by a small coil located above the cell and aligned along the $\hat{x}$ axis. The magnitude and phase of the atomic coherence is mapped onto the polarisation of the linearly polarised probe beam through the paramagnetic Faraday effect \cite{Takahashi1999, Savukov2005}. The probe laser frequency can be tuned $+\SI{7}{\giga\hertz}$ to $-\SI{2}{\giga\hertz}$ from the $6\,^2$S$_{1/2}$ F=3 $\rightarrow{}6\,^2$P$_{3/2}$ F'=2 transition via phase-offset-locking to the pump beam, and propagates orthogonally to the pump. The rf resonance signal is measured by a lock-in amplifier referenced to the first harmonic of the rf field frequency.   

\begin{figure}[h!]
\includegraphics[width=\columnwidth]{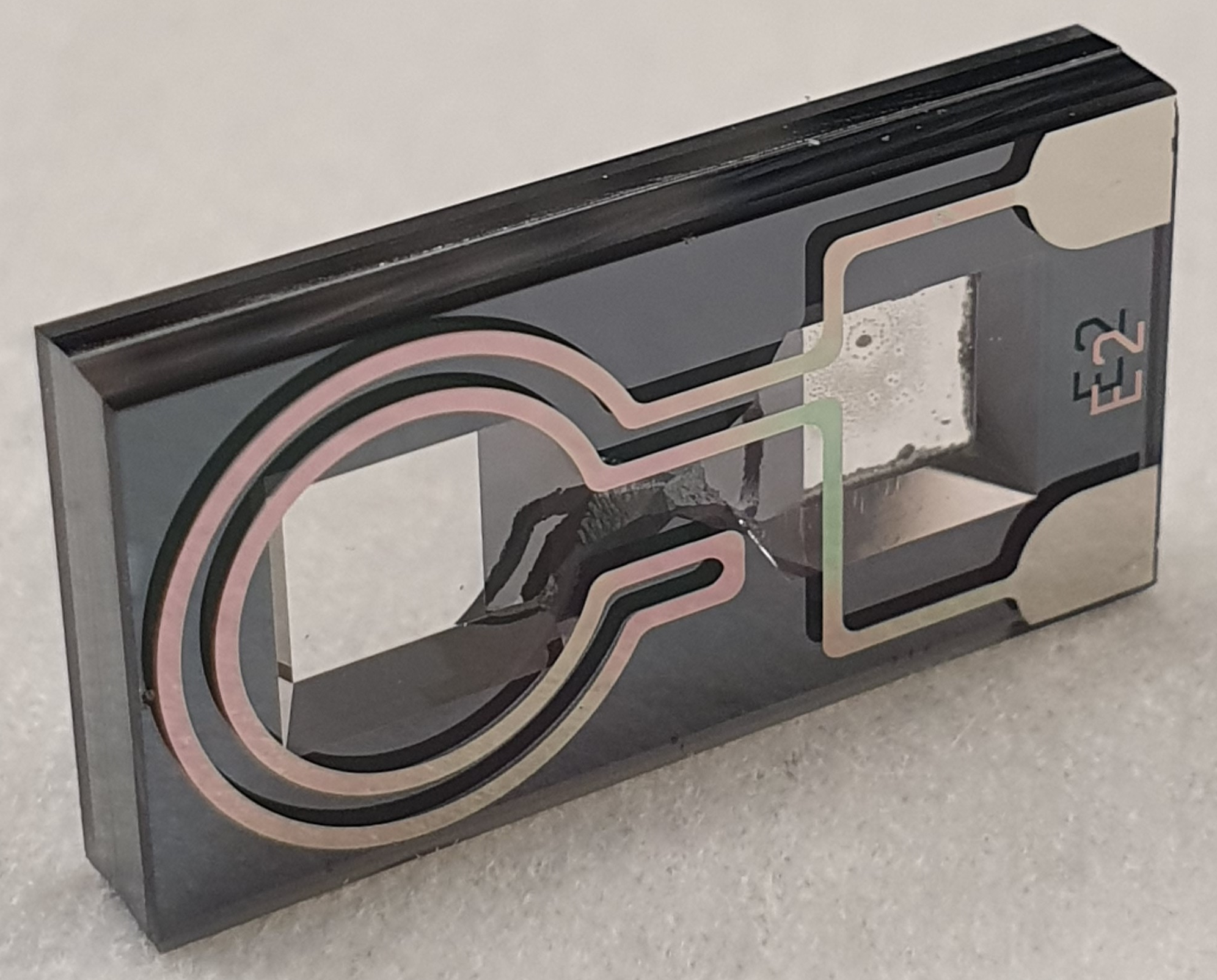}
\caption{
 Silicon-glass wafer fabricated vapour cell with platinum circuits on both windows, forming the heating unit. On the right end side of the cell it is visible the storage chamber containing the caesium sample.
}\label{fig:Cell}
\end{figure}

\section{Silicon wafer-based cell}

\subsection{Layout and manufacturing}

As previously mentioned, the wafer-based vapour cell used in this work has a double-chamber design,  Fig.~\ref{fig:Cell}. One chamber is used as storage for the alkali precursor, caesium azide, and it is connected to a second, the interaction chamber, where the released $Cs$ vapour interacts with the laser beams. The vapour cell was constructed using an anodic bonded triple stack of $BF33$ glass/$Si$/$BF33$ glass. The $Si$ layer is $\SI{2}{\milli\meter}$ thick, and the wafer has a diameter of $\SI{150}{\milli\meter}$. The chambers and channels were etched using a $Si_3N_4$ hard mask and a $KOH$ solution. A $\SI{0.5}{\milli\meter}$ thick glass wafer was bonded on one side of the $Si$ to create an open-top chamber. Caesium azide in deionised water solution was dispensed into the storage chamber and then dried. A second $\SI{0.5}{\milli\meter}$ thick glass wafer was anodically bonded to the $Si$ to create a sealed cell. Before sealing the chambers, it is possible either to evacuate to ultra-high vacuum or to add an additional buffer gas. After bonding, the caesium azide was decomposed using broadband UV radiation, resulting in the release of $Cs$ and $N_2$. The cells were extracted from the wafer stack using a dicing saw. The cells have dimensions of  $3 \times 20.2 \times \SI{10.2}{\milli\meter}^3 $, with an interaction chamber of $2 \times 4 \times \SI{4}{\milli\meter}^3 $, and weight of $1.24$ g. 

On-chip heating was fabricated onto the vapour cell to reduce the overall size, decrease power consumption, ease integration, and generate a thermal gradient to inhibit the formation of $Cs$ condensation on the windows of optical access. COMSOL simulations, described below, were used to design a $Pt$ heating circuit with minimal stray magnetic field. The design was then transferred to a lithographic mask, which greatly simplified the fabrication process. Standard wafer lift-off resist processing and e-beam evaporation were used to define the heater tracks and pads. The on-chip heaters were passivated using SU-8 photoresist, hard-baked to make a permanent layer over the $Pt$ tracks.

\subsection{Modelling}

Modelling and design of the sensor head were carried out using the finite element (FE) software package COMSOL Multiphysics, which allowed us to calculate both magnetic field and temperature maps of the cell. 

\begin{figure}[h!]
\includegraphics[width=\columnwidth]{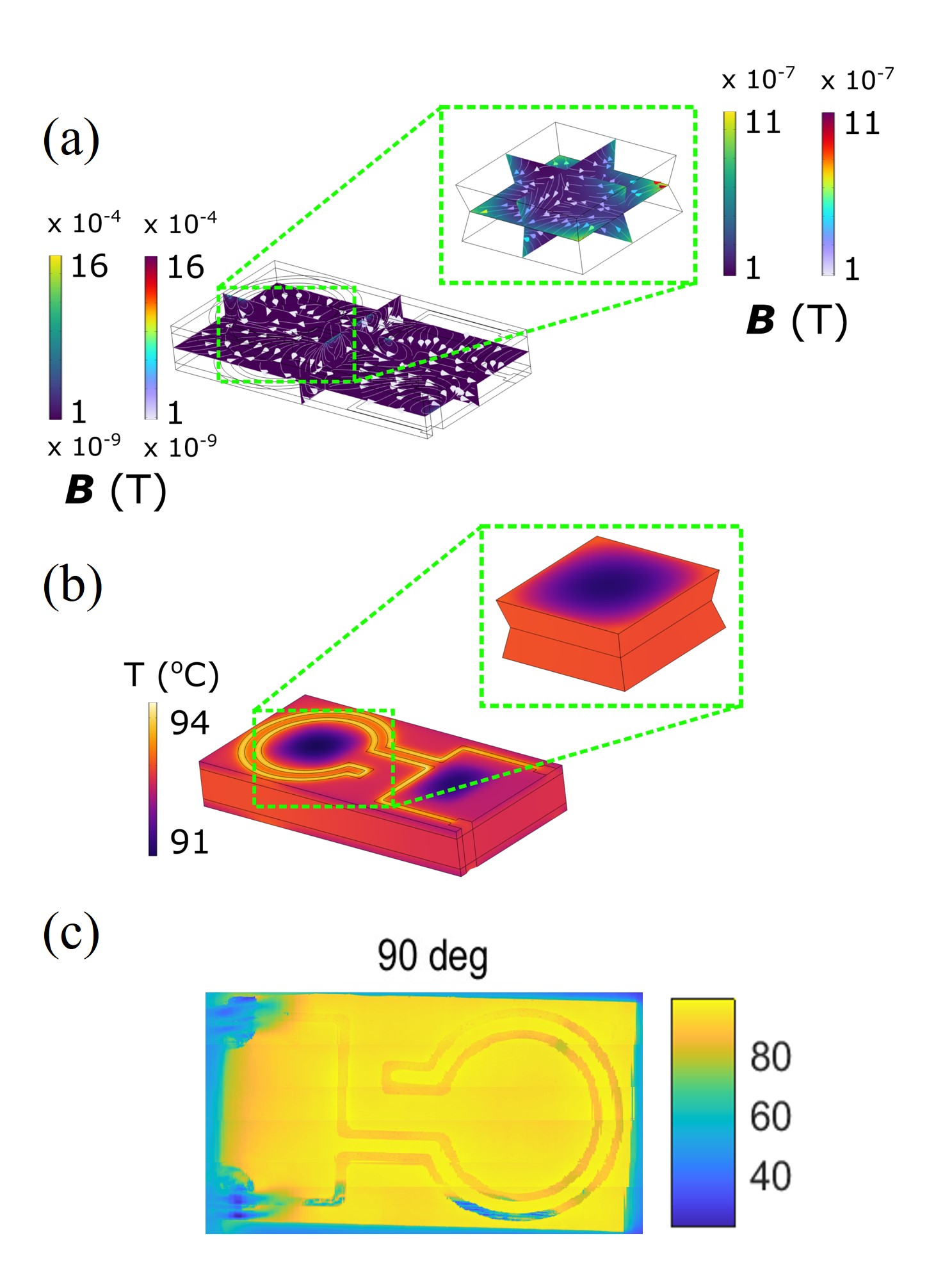}
\caption{
 (a) Modelled magnetic flux density \textbf{B} through the device, as generated by the heater with DC current input. (b) Modelled surface temperature of the cell for an input voltage of 10.5 V. The hottest regions of the model are the platinum heaters, whilst the coldest regions are the regions above the chambers. The surface temperature of the interaction chamber is shown inset. (c)  Surface cell temperature map measured through thermal imaging. Heater tracks appear darker due to platinum’s lower emissivity.
}\label{fig:Model}
\end{figure}

The shape and dimensions of the heating elements were designed to minimise the magnetic flux generated by the resistive heating and to maintain a stable temperature across the interaction chamber without blocking the optical access. The model also includes the copper pin contacts of $\SI {1.3}{\milli\meter}$ wide, $\SI {18}{\milli\meter}$  long, $\SI {0.15}{\milli\meter}$ thick to more accurately model the temperature. The heater is configured such that the IN/OUT current wires encircle the interaction chamber, and are arranged in close proximity, to minimise the generated magnetic field. Fig.~\ref{fig:Model} (a) shows the magnetic flux density \textbf{B}, generated by the heating elements with DC current input, through the device and through the interaction chamber. It is worth noting that even though \textbf{B} is quite large in some parts of the device, and could thus impact the atoms contained in the chambers, the external correction coils and AC current input effectively help to minimize this issue. In Fig.~\ref{fig:Model} (b) we show a surface map of the temperature obtained from the model with an input voltage of 10.5V (the current used in the FE models was DC due to the limitations of the software used). As one would expect, the hottest regions correspond to the heaters themselves, with a maximum temperature of $78.5^{\circ}C$ for this input voltage. The coldest regions in the device are found in the centre of the glass windows, as a consequence of the heater design. Figure~\ref{fig:Model} (b) also shows a zoomed-in view of the surface temperature in the interaction chamber, which has a $1.59^{\circ}C$ total temperature differential across its extension. 

\begin{figure}[h!]
\includegraphics[width=\columnwidth]{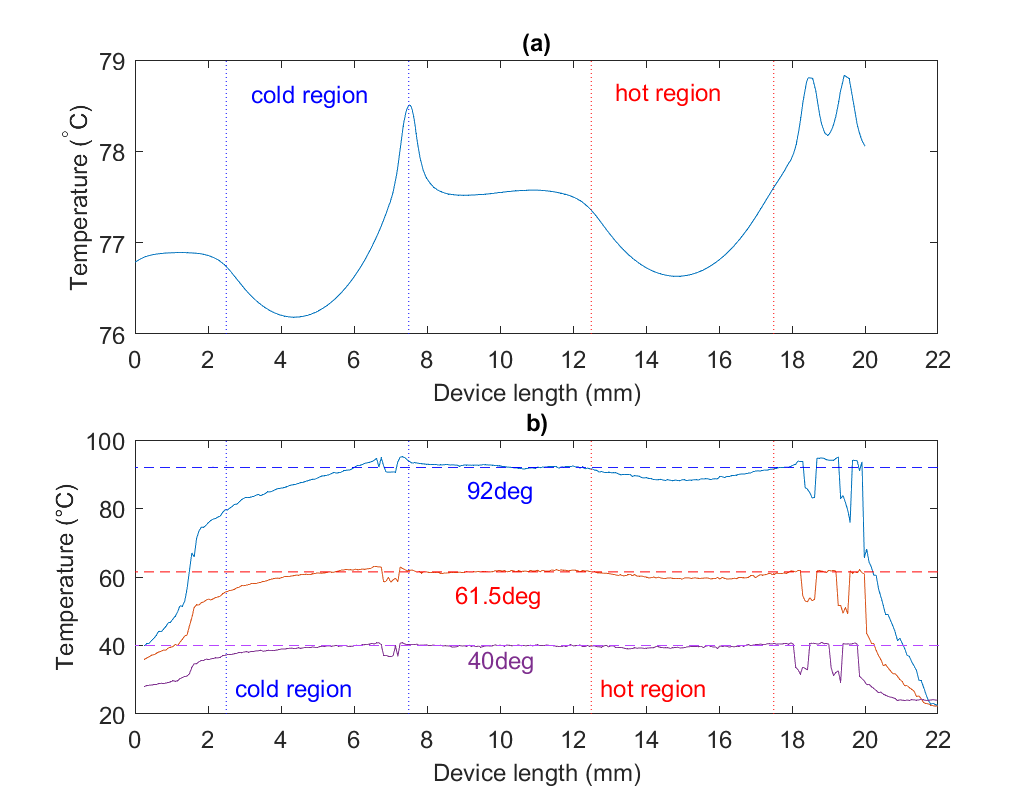}
\caption{
Modelled (a) and measured (b) temperature line trace across the upper surface of the device, where the interaction (red) and storage (blue) chamber regions are highlighted. 
}\label{fig:Temperature}
\end{figure}

Figure~\ref{fig:Temperature} (a) shows a temperature line-trace across the upper surface of the device for an input voltage of 10.5 V.  The interaction and storage chambers are indicated by the blue and red regions respectively. The temperature drops across both areas, as one would expect given the cold spots shown in Fig.~\ref{fig:Temperature} (a), however, there is a temperature gradient between these two chambers which ensures that the caesium will condensate predominantly in the storage chamber. The three peaks at roughly $78.5^{\circ}C$ correspond to the $Pt$ wires. 


\subsection{Charaterisation}
We characterised the temperature distribution across the cell by using an Infratec ImageIR 8300 thermal imager with a 3.0x M90175 lens, pixel size of $ \SI {5}{\micro\meter}$, and a $ \SI {466}{\micro\second}$ integration time to measure  cell's emissivity. The cell was painted using black spray paint to limit the effect of relative emissivity differences between materials (glass, silicon, platinum). The heaters were wired in series to a DC power supply unit, with a total resistance of $\SI {573}{\ohm}$ at room temperature. The cell was mounted to a two-axis motorised translation stage to enable systematic movement. The cell was then rastered beneath the imager, and the images were subsequently combined together. An example of a combined image is given in Fig.~\ref{fig:Model} (c), showing the temperature distribution across the cell with a temperature in the centre of the cell of $92^{\circ}C$. Figure~\ref{fig:Temperature} (b) shows a temperature line-trace across the centre of the upper surface of the device for three different maximum temperatures (92, 61.5, and 40 $^{\circ}C$). The results confirm that the interaction chamber (hotter region) has a higher average temperature than the storage chamber (colder region) in all cases. This temperature gradient is greatly beneficial for avoiding condensation of \textit{Cs }in the interaction chamber, especially along the optical access. A higher temperature gradient across the cell between chambers is observed for increasing temperature. This is partially due to the cell mounting and electrical contacts, which act as heat sinks. It should also be noted that, in these measurements, the $Pt$ heating wires appear to be at a lower temperature than the surroundings, which is due to the emissivity differences between the $Si$ (high) and $Pt$ (low). While this could be avoided with additional sample preparation steps, we consider it sufficient for the purpose of the present characterisation.  

\section{Dynamics of the polarised atomic vapour}

\subsection{Indirect pumping}
The signal obtained from a magnetometer and, more generally, the dynamics of the atomic polarisation and coherence generation, are determined by the imbalance between the atomic polarisation/coherence creation rate and the various decoherence rates. There are four factors that mainly contribute to these rates: wall collision, diffusion, optical excitation, and SEC. As it will be discussed below, optical excitation, with a pumping rate defined by the pump beam power, and SEC, with a rate defined by the atomic vapour temperature, contribute both to the signal (coherence) generation and decoherence. Optical pumping produces stationary collective atomic spin components and consequently contributes to coherence generation. However, in a regime where the pumping rate is larger than the atomic coherence generation rate, it contributes to additional decoherence. In turn, SEC introduce couplings between atoms in the form of exchange of polarisation, which could contribute either to dephasing or to transfer of spin temperature, \cite{Chalupczak2012a} and coherence \cite{Ruff1965, Haroche1970, Skalla1996, Chalupczak2014}. 
The term 'indirect pumping' refers to a process of generation of atomic polarisation that combines optical excitation and SEC \cite{Chalupczak2012a}. The process is analogous to the so-called 'spin-exchange optical pumping' that involves transfer of polarisation between optically pumped alkali-metal and noble gas atoms \cite{Walker1997}. Indirect pumping has been demonstrated in paraffin-coated cells \cite{Gartman2015}. In this case, tuning the pump laser frequency close to the $6\,^2$S$_{1/2}$ F=3 $\rightarrow{} 6\,^2$P$_{3/2}$ F'=2 transition (for caesium) ensures that polarisation within the F=3 manifold is generated via direct optical pumping, while atoms in the F=4 manifold are not directly coupled to light. The transfer of polarisation from the F=3 to the F=4 ground state level is thus realised only via spin-exchange collisions. Eventually, when the pump becomes strong enough to broaden the transition, atoms are also transferred from the F=3 to the F=4 ground state \cite{Chalupczak2012a}. The combination of indirect pumping, optical transfer, and SEC produces an accumulation of atoms in the stretched state of the F=4 level. Conservation of momentum in the SEC process results in immunity of the F=4 stretched state to SEC decoherence, which becomes visible as a narrowing of the rf resonance linewidth \cite{Chalupczak2012a}. In fact, the linewidth of the rf spectra produced by low polarized atomic ensemble at Larmor frequencies below $\sim \SI {200}{\kilo\hertz}$ benefits from the narrowing resulting from the degeneracy of all F=4 coherence frequencies and consequently, coherence transfer within the F=4 level \cite{Chalupczak2014}. In other words, due to the degeneracy of all F=4 coherence frequencies only SEC that cause the F=4 to F=3 spin flips contribute to decoherence.
Note also that, in paraffin-coated cells, the physical shielding of the atoms from the cell walls allows the atoms to acquire global properties over an integration time set by the fast ballistic dynamics. As a result, all atoms can similarly interact with the pump beam, and between themselves through SEC.

\begin{figure}[h!]
\includegraphics[width=\columnwidth]{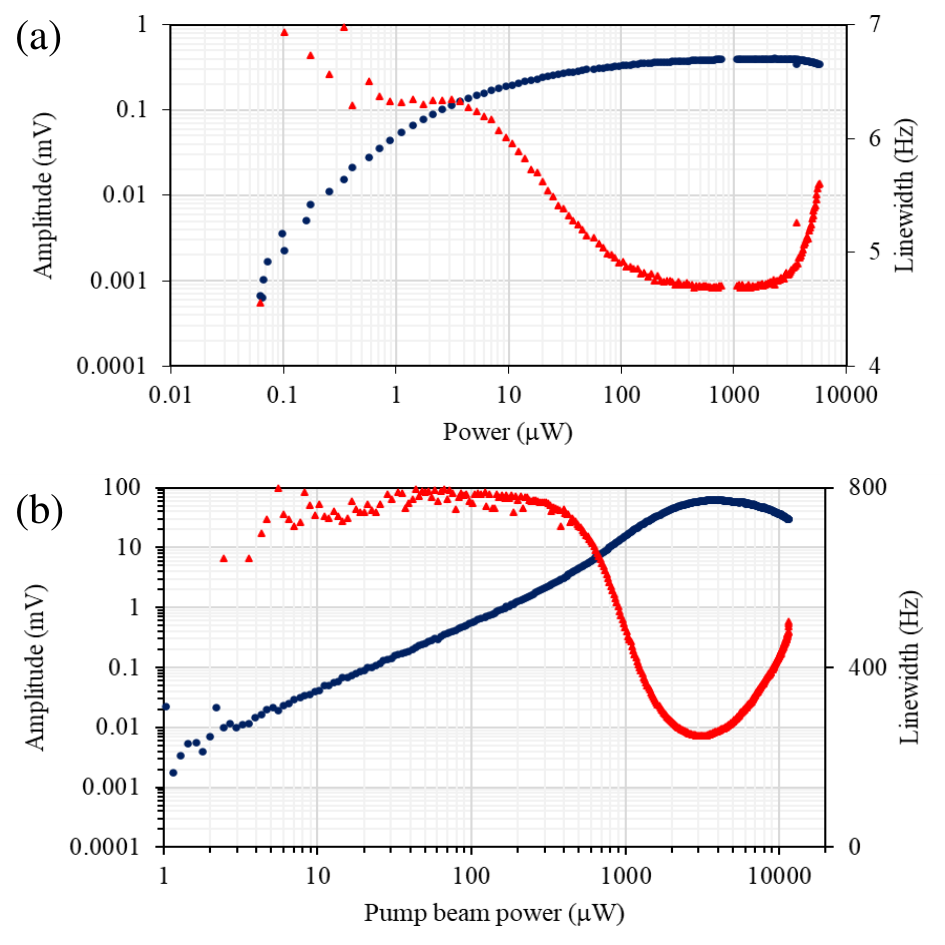}
\caption{
Amplitude (blue dots), and linewidth (red triangles) of the rf spectral profile as a function of the pump beam power recorded in room-temperature paraffin-coated (a) and wafer-based cells (b) at temperature of $\sim 110^{\circ}$C. The Larmor frequency for both measurements is $\SI {15}{\kilo\hertz}$. 
}\label{fig:Pump_power_scan}
\end{figure}

\subsection{Measurements}
To gain insight into the role of optical excitation and SEC, the dependence of the rf spectroscopy signal on the pump beam power is studied for various measurement conditions. The rf spectrum is created by monitoring the magneto-optical signal as the rf field frequency is scanned across the magnetic resonance \cite{Chalupczak2018}. The rf resonance amplitude provides a measurement of the magnitude of the F=4 atomic coherence, while its linewidth is related to decoherence processes. The resonance frequency, linewidth, and amplitude of the spectral profile have been extracted by fitting a single Lorentzian, or a combination of two Lorentzian profiles to the rf spectrum (see discussion below).

In particular, we compare the rf spectroscopy signals obtained in a paraffin-coated cell at room temperature with those in a wafer-based cell at a vapour temperature $\sim 110^{\circ}$C with the same optical pumping on the above transition, Fig.~\ref{fig:Pump_power_scan}. Typical pump beam power scan data recorded in paraffin-coated cells, Fig.~\ref{fig:Pump_power_scan} (a), show an increase of the signal amplitude accompanied by the narrowing of the spectral profile, up to roughly $\SI{200}{\micro\watt}$. This narrowing results from the combination of effects described in the previous paragraph \cite{Chalupczak2012a}. At low powers ($< $$\SI{10}{\micro\watt}$), optical pumping on the closed transition generates a population imbalance within the F=3 level, which is collisionally transferred to the F=4 level via SEC. For intermediate pump powers ($\SI{10}{\micro\watt}$ - $\SI{1}{\milli\watt}$), the addition of optical transfer from F=3 to F=4 together with SEC recycling between the two hyperfine ground states effectively favours the occupation of the F=4 stretched state. Note that this process is different from the so-called 'light narrowing', which is caused by the suppression of the spin-exchange broadening at high spin polarisation created by high power laser \cite{ Appelt1999}. Finally, for pump beam powers above $\SI{1}{\milli\watt}$, decrease of the signal and broadening of the profile is observed. This is a consequence of the optical pumping rate being larger than the SEC rate, rather than off-resonant optical coupling to the F=4 atomic level \cite{Chalupczak2013}. 

We observe that in some parameter subspace the signal amplitude (blue dots) and linewidth (red triangles) measured in the wafer cell show a similar dependence on the pump power as in the paraffin-coated one. This suggests that an analogous narrowing mechanism might take place in the two cells, despite the very different experimental conditions. In detail, the data recorded in the wafer-based cell show that the linewidth of the rf resonances (red triangles in Fig.~\ref{fig:Pump_power_scan} (b)) for pump powers below $\SI{300}{\micro\watt}$ does not change significantly with the pump beam power. The data set in Fig.~\ref{fig:Pump_power_scan} (b) (blue dots) shows also that the amplitude of the F=4 component increases linearly with the pump power up to $\SI{1}{\milli\watt}$. In principle, and differently from the case of a paraffin-coated cell, optical pumping creates orientation within the F=3 level and transfers atoms to the F=4 state already at low pump powers, due to collisional broadening from the buffer gas. However, it is going to shown below that this mechanism does not always work well in practise, and that the linewidth value at low pump powers is still mainly limited by SEC in this regime of parameters. In wafer cells with a buffer gas, atomic diffusion combined with the boundary conditions set by the cell walls, results in the generation of spatial modes that retain local, instead of global, properties of the cell, and generally show distinct magnetic features \cite{Shaham2020}. Different spatial modes manifest themselves through distinguishable components in the rf spectra. This allows, for example, the signal related to the lowest order diffusive mode to be singled out.

At a temperature of 110 $^{\circ}$C, where the atomic sample is optically thick, only a fraction of the atoms in the extended, lowest diffusion mode directly interacts with the pump light, even though the beam had a homogeneous radial profile, i.e. a ‘top hat’ profile. As a result, the effect of the pump is hampered. In a pump power range between $\SI{300}{\micro\watt}$ and $\SI{6}{\milli\watt}$ narrowing of the profile is observed. There are two mechanisms responsible for this process: increasing the optical pumping rate and transfer, and an increasingly more uniform optical pumping. In these conditions, the cooperative action of SEC selective decoherence and optical excitation again leads to generation of the F=4 orientation, with large occupation of the stretched state, similar to the case already described of a paraffin-coated cell. 
Approaching the optical saturation of the cell, $\SI{3}{\milli\watt}$, the signal amplitude reaches a maximum and then gradually decreases. This is accompanied by a progressive broadening of the rf spectrum profile. As it will be shown below, this behaviour has a common pattern for the pump power dependencies recorded across the whole temperature range of our measurements ($25-120^{\circ}$C). 

\begin{figure}[h!]
\includegraphics[width=\columnwidth]{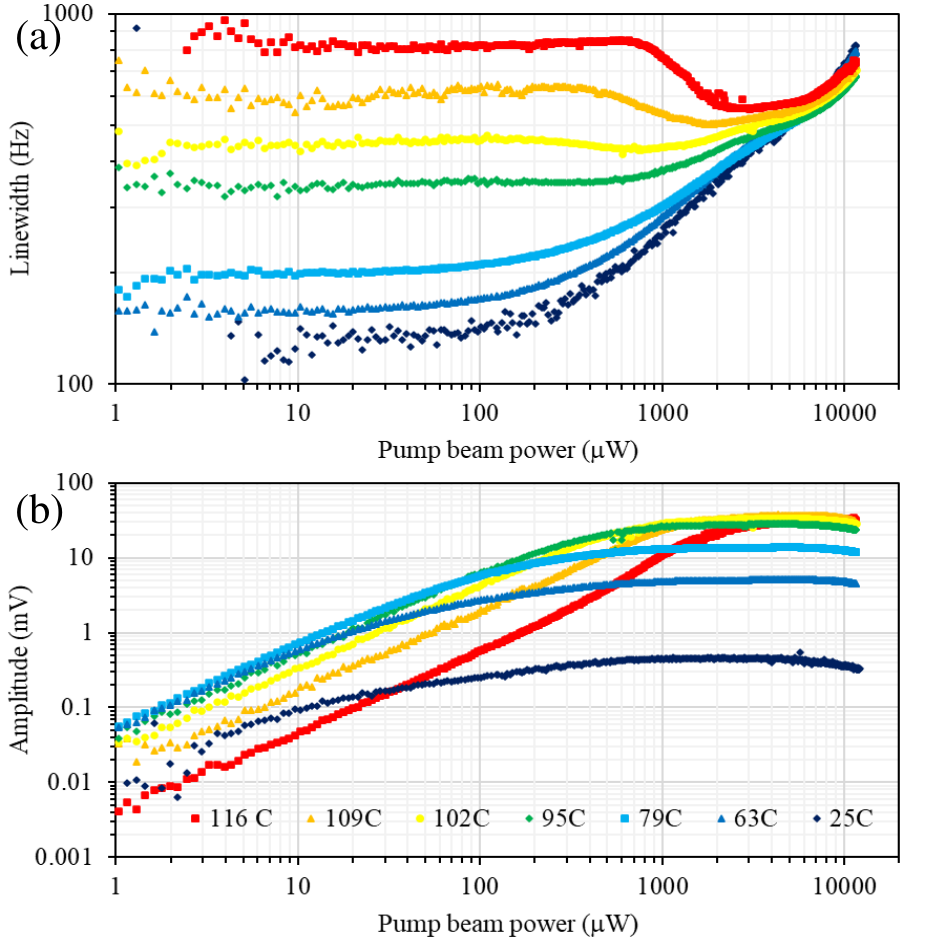}
\caption{Linewidth (a) and the amplitude (b) of the rf spectral profile as a function of pump beam power recorded at temperatures ranging between $25^{\circ}$C and $\sim 120^{\circ}$C. The power dependencies of the rf spectral profile are recorded for the Larmor frequency of $\SI{5}{\kilo\hertz}$.
}\label{fig:Spectra}
\end{figure}

\subsection{Temperature dependence} 
\begin{figure}[t!]
\centering
\includegraphics[width=\columnwidth]{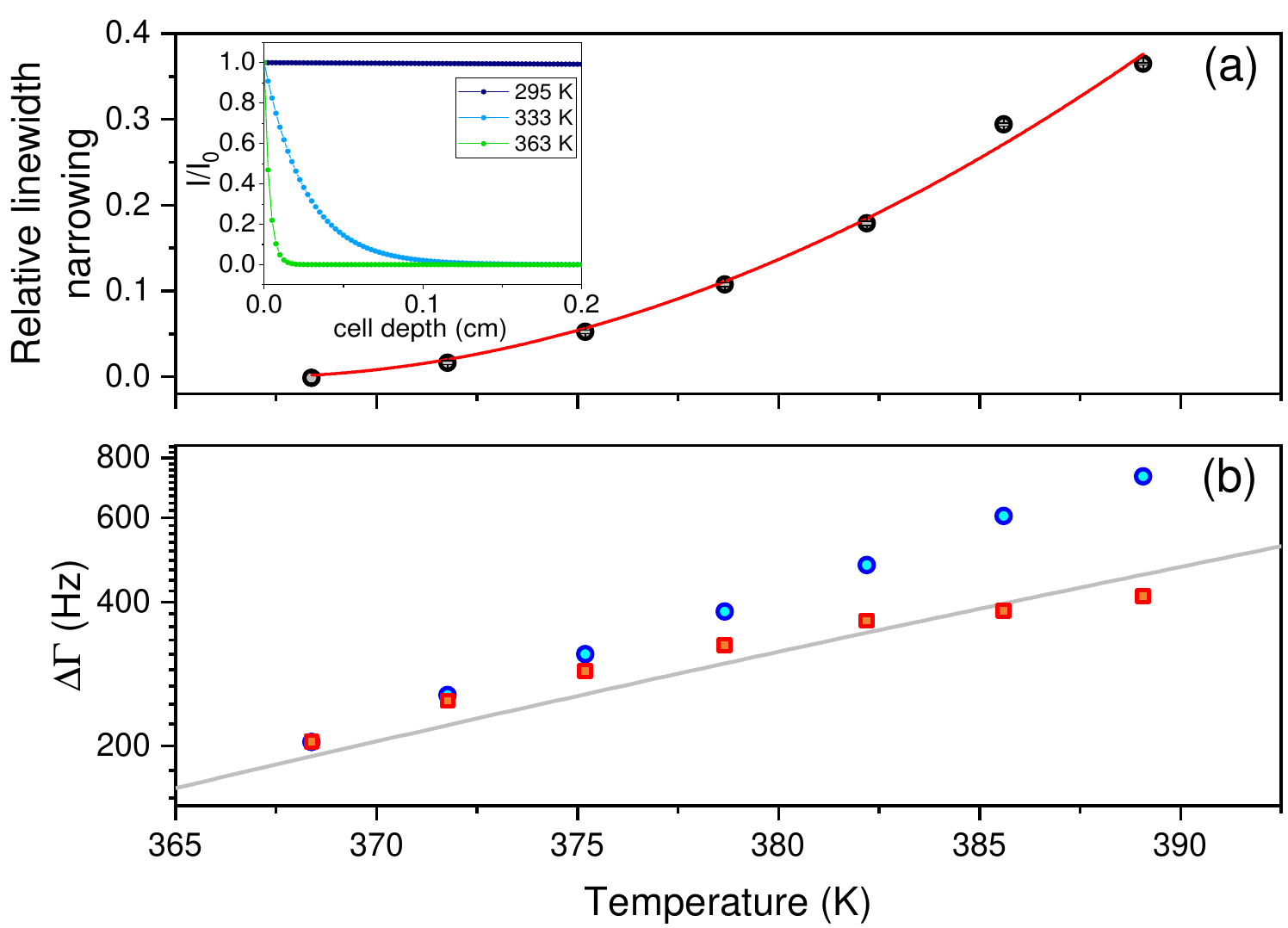}
\caption{(a) The relative line narrowing measured as a function of temperature. The inset shows the calculated transmittance of the pump beam across the cell, for different temperatures and a fixed pump power of $\SI{1000}{\micro\watt}$. (b) The increase in linewidth (with respect to room temperature measurements) as a function of temperature (blue dots) measured between pump powers of $\SI{100} {\micro\watt}$-$\SI{200} {\micro\watt}$, and in correspondence to the maximum line narrowing (red squares). The gray solid line shows the extrapolation from a fit of the temperature dependence below $90^{\circ}$C. 
}
\label{fig:Temperature2}
\end{figure}

In order to show the role of the optical excitation and SEC in the rf signal generation in wafer-based buffer cell, we record scans showing the pump power dependence for atomic vapour temperatures in the range $25-120^{\circ}$C, Fig.~\ref{fig:Spectra} (a)-(b). Recording the whole pump power dependence of rf spectra in contrast to a single spectrum at specific conditions allows us to better identify the different mechanisms contributing to the signal behaviour. 
No significant change in the linewidth values is observed with the vapour temperatures below $40^{\circ}$C. This indicates that collisions with the cell walls and the buffer gas dominate over SEC decoherence in the temperature range $25^{\circ}$C-$40^{\circ}$C. Also, the increase of the linewidths observed above $\SI{100}{\micro\watt}$ reveals the decoherence introduced by the pump beam. This light-induced decoherence rate is common to all the temperature data sets in the high pump power regime. SEC contribution to decoherence instead grows with increasing vapour temperature, and hence density, at a rate of roughly $\SI {40}{\hertz}$/$^{\circ}$C. The signal linewidth gradually becomes temperature-dependent,  Fig.~\ref{fig:Spectra} (a). Signal profile narrowing then appears for temperatures above $90^{\circ}$C, at pump powers above a few $\SI{100}{\micro\watt}$. Notably, the relative narrowing (i.e. the ratio between the narrowing and the linewidth measured around $\SI{100}{\micro\watt}$ pump power) scales roughly quadratically with the temperature (exponent equal to $1.9(1)$),  Fig.~\ref{fig:Temperature2}(a). Moreover, the rate at which the linewidth grows with temperature above $90^{\circ}$C is higher than for lower temperatures. This can be seen in Fig.~\ref{fig:Temperature2}(b), where the increase in linewidth (blue dots) is shown together with the functional dependence extrapolated from a fit to the data between $40-90^{\circ}$C (grey solid line). This suggests that not only is the SEC rate increased with respect to the optical pumping rate (for a given pump power), but also that an additional mechanism is responsible for hampering the pump for higher atomic densities. In fact, a theoretical estimate shows that by increasing the atomic density for a given pump power, the atomic sample changes from being optically thin to thick, inset of Fig.~\ref{fig:Temperature2} (a). Differently from the case of paraffin coated cells, in buffer cells, where the atomic gas diffuse relatively slowly with respect to the rf-induced dynamics and gets depolarized at the cell walls, local conditions in the cell (such as those set by the size and position of laser beams) matter. Moreover, the fact that the linewidth does not significantly vary with temperature for powers larger than $2-\SI{3}{\milli\watt}$, indicates a high level of global polarisation, accompanied by transfer into the stretched state of the F=4 level.   

Spectral profile narrowing is also accompanied by an increase in the signal amplitude. This is seen as a departure from the linear pump power dependence evident in Fig.~\ref{fig:Spectra} (b), and also visible in Fig.~\ref{fig:Pump_power_scan} (b). The same trends as presented in Fig.~\ref{fig:Spectra}, were observed in a different spherical buffer gas glass cell with $\SI{20}{\milli\meter}$ diameter, and a buffer gas made of 300 Torr Neon, 50 Torr of Nitrogen. From the comparison with this cell, we can also individuate the different contributions to the linewidth in a regime of low pump power and low temperature, due to the different diffusion coefficient (which depends on the buffer gas pressure), and size of the cells \cite{Vanier1974}. Finally, for a temperature of $\sim 120^{\circ}$C the maximum signal amplitude observed in the pump power scans starts decreasing below the level obtained at slightly lower temperatures. This is a consequence of the optical density becoming significant enough to prevent the atomic cloud from saturating in the range of laser power available. 
The trade-off of these parameters is relevant for optimization of the performance of an atomic magnetometer. For instance, optical density larger than 1 guarantees that the sensor can truly operate in a relevant quantum regime, i.e. that the atomic projection noise dominates over the photon shot noise \cite{Kuzmich1998}. Noise spectra recorded above $90^{\circ}$C fall in this regime where atomic projection noise and back action dominate over photonic shot noise.  


\begin{figure}[h!]
\includegraphics[width=\columnwidth]{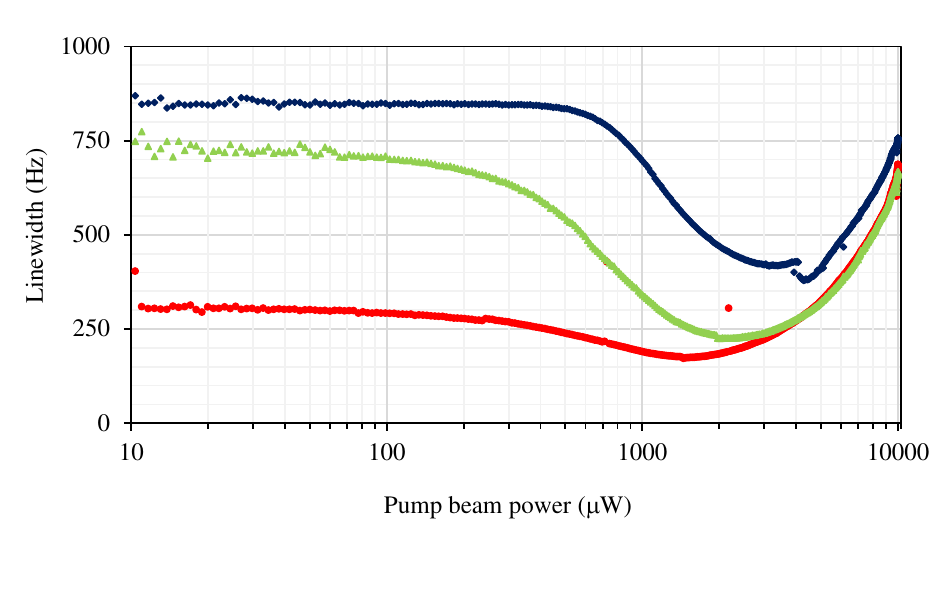}
\caption{ Linewidth of the rf spectral profile as a function of the pump beam power recorded at Larmor frequencies $\SI{\sim 120}{\hertz}$ (red dots), $\SI{570}{\hertz}$ (green triangles), $\SI{3.2}{\kilo\hertz}$ (blue diamonds) in a vapour cell temperature of $120^{\circ}$C.
}\label{fig:Larmor}
\end{figure}

\subsection{Magnetic field dependence}

As stated above, the SEC-induced decoherence detected in the rf profiles linewidth recorded at low pump beam powers is due to spin flips between F=4 and F=3. This is confirmed by recording the dependence of the pump powers scans on the bias magnetic field strength, in a high temperature regime ($>90^{\circ}C$). Changes in the linewidth of the rf profiles, recorded at low pump powers, with bias field strength are mainly the signature of the coherence transfer between the F=4 and F=3. 

Although the F=3 and F=4 atomic coherences oscillate with opposite frequencies, for spectral linewidths exceeding the Larmor frequency, these frequencies become indistinguishable, which allows the macroscopic transfer of the atomic coherences between the levels without phase mismatch. This phenomenon is sometimes referred to as the Spin-Exchange Relaxation Free regime \cite{Happer1973, Happer1977, Appelt1998, Walter2002, Kominis2003}. Figure ~\ref{fig:Larmor} shows the power dependence of the rf spectral profile linewidth recorded at Larmor frequencies of $\SI{\sim 120}{\hertz}$ (red dots), $\SI{570}{\hertz}$ (green triangles), $\SI{3.2}{\kilo\hertz}$ (blue diamonds) in a vapour cell temperature of $120^{\circ}$C. 
These data series show a few generic trends that were observed also in the spherical buffer glass cells. Firstly, the values of the linewidth decreases with decreasing Larmor frequency, with the linewidths recorded at low pump power showing the biggest relative change. Then the linewidth narrowing recorded for pump powers larger than $\SI{1}{\milli\watt}$ becomes shallower. This reflects the decreasing significance of pumping to the stretched state with the decrease of SEC-induced decoherence. 
Finally, it was noticed that the decrease in linewidth, generally results in an increase in the signal amplitude, across a range of the pump powers. The value of the pump power for which the signal maximum is observed decreases with decreasing Larmor frequency. A reduction in the relative light shift was also observed for the pump powers below $\SI{3}{\milli\watt}$. 
The narrowest linewidth recorded at $112^{\circ}$C is $\SI{95}{\hertz}$. In similar conditions, we were able to observe linewidths as narrow as $\SI{8}{\hertz}$ in the spherical buffer gas cell, which is soley due to the difference in diffusion coefficient between the two cells.

\section{Conclusions}
We presented the design and characterization of miniature silicon wafer-based cells, with double-chamber configuration and an attached $Pt$ circuit for heating. Using finite element modelling, the thermal and magnetic properties of the cells' design have been highlighted in detail. The novel cells have been extensively characterized by systematically studying the dynamics of the atomic vapour polarisation in different conditions (pump power, temperature, magnetic field bias). The contributions of indirect optical pumping and SEC were discussed, and a comparison with a state-of-the-art reference paraffin-coated cell was drawn. This allowed us to find a qualitatively similar behaviour in the two cells in different regimes, and to identify a similar light-induced line narrowing mechanism. It needs to be stressed that indirect pumping is crucial for the observation of such narrowing, which would be otherwise concealed by pump-induced power broadening.

Metal condensation on cell windows may compromise magnetometer performance and reproducibility of the results. In early stages of the buffer cell development, it was noticed that: \begin{quote}
    ‘These small droplets do not interfere with the optical pumping and they help to ensure that the alkali-metal-atom vapor pressure is close to the saturated vapor pressure at the cell temperature’.  \cite{zeng85}
\end{quote} 
In series measurements with buffer gas glass cells, we have confirmed that the presence of these droplets defines the relaxation rate of the atomic coherence and effectively sensitivity of the atomic sensor. Changing dimensions of the droplets observed when the cell temperature is varied results in limited reproducibility of the observed rf profiles linewidth. While we have observed variation in decoherence rate up to a level of 30\% in glass cells, this level is reduced to 5\% in wafer cells

In conclusion, we have presented a robust, simple, and inexpensive realization of an atomic magnetometry architecture, with excellent performances for non-zero field operation, and for applications requiring miniaturisation. 
While the focus of the present work has been on the discussion of processes involving the main diffusion atomic mode, the manipulation of higher order spatial modes, and of intermode coupling \cite{Bevington23} will be a topic for future developments. This will open opportunities to use our wafer-based cells not only for portable sensors, but also as a tool for exploration of fundamental aspects of the atomic vapour dynamics, and for further applications in the realm of quantum imaging and information. Indeed, thanks also to the maturity of the manufacturing process, wafer-based cells are very appealing substrates for the development of atom-photon interfaces \cite{Shaham2022}. Integrating cells with other wafer-based photonic devices (such as laser sources and detection units), and miniaturized components (such as magnetic coils and shields), would open the way to the realization of cm-sized fully-equipped atomic units of great interest for several Quantum Technology applications.

\section{Acknowledgment}
 
We acknowledge the support of the UK government Department for Science, Innovation and Technology through the UK national quantum technologies program. We would like to thank A. Parsons for critically reading this manuscript, and J. McMillan for their help with thermal imaging. 

\section{Data Availability Statement}
The data that support the findings of this study are available from the corresponding author upon reasonable request.

\end{document}